\documentclass[superscriptaddress,prl,twocolumn,nofootinbib]{revtex4}
\usepackage{amsmath,amssymb}
\usepackage{graphicx,epsfig}
\usepackage{slashed}
\usepackage{float,xcolor}

\def\beq{\begin{equation}}
\def\eeq#1{\label{#1}\end{equation}}
\def\eeqn{\end{equation}}
\def\beqa{\begin{eqnarray}}
\def\eeqa#1{\label{#1}\end{eqnarray}}
\def\eeqan{\end{eqnarray}}
\def\CR{\nonumber \\ }
\def\leqn#1{(\ref{#1})}

\def\mev{\mathrm{\ MeV}}

\newcommand{\centeron}[2]{{\setbox0=\hbox{#1}\setbox1=\hbox{#2}\ifdim
\wd1>\wd0\kern.5\wd1\kern-.5\wd0\fi \copy0
\kern-.5\wd0\kern-.5\wd1\copy1\ifdim\wd0>\wd1
                                   \kern.5\wd0\kern-.5\wd1\fi}}
\newcommand{\ltap}{\>\centeron{\raise.35ex\hbox{$<$}}
                           {\lower.65ex\hbox{$\sim$}}\>}
\newcommand{\gtap}{\>\centeron{\raise.35ex\hbox{$>$}}
                           {\lower.65ex\hbox{$\sim$}}\>}
\newcommand{\gsim}{\mathrel{\gtap}}

\def\gap{M_{\rm gap}}
\def\rhoc{\rho_{\scriptstyle_{\rm CFT}}}
\def\Pc{P_{\scriptstyle_{\rm CFT}}}
\def\rhos{\rho_{\scriptstyle_{\rm SM}}}
\def\Ps{P_{\scriptstyle_{\rm SM}}}
\def\l{\lambda_{\scriptstyle_{\rm CFT}}}
\def\L{\Lambda_{\scriptstyle_{\rm CFT}}}
\def\Oc{{\cal O}_{\scriptstyle_{\rm CFT}}}
\def\Os{{\cal O}_{\scriptstyle_{\rm SM}}}
\def\Tc{T_D}
\def\mdm{m_{\scriptstyle_{\rm DM}}}
\def\ds{d_{\scriptstyle_{\rm SM}}}

\newcommand{\CFT}{\mathrm{CFT}}
\newcommand{\SM}{\mathrm{SM}}
\renewcommand{\d}{\mathrm{d}}

\begin{document}

\title{Conformal Freeze-In of Dark Matter}

\author{Sungwoo Hong}
\email{sungwoo.hong@cornell.edu}
\author{Gowri Kurup}
\email{gk377@cornell.edu}
\author{Maxim Perelstein}
\email{mp325@cornell.edu}
\affiliation{Laboratory for Elementary Particle Physics, Cornell University,
Ithaca, NY 14850, USA}

\date{\today}

\begin{abstract}
We present the conformal freeze-in (COFI) scenario for dark matter production. At high energies, the dark sector is described by a gauge theory flowing towards a Banks-Zaks fixed point, coupled to the standard model via a non-renormalizable portal interaction. In the early universe, a non-thermal freeze-in process transfers energy from the standard model plasma to the dark sector. During the freeze-in, the dark sector is described by a strongly coupled conformal field theory. As the universe cools, cosmological phase transitions in the standard model sector, either electroweak or QCD, induce conformal symmetry breaking and confinement in the dark sector. One of the resulting dark bound states is stable on the cosmological time scales and plays the role of dark matter. With the Higgs portal, the COFI scenario provides a viable dark matter candidate with mass in a phenomenologically interesting sub-MeV range. With the quark portal, a dark matter candidate with mass around $1$ keV is consistent with observations. Conformal bootstrap may put a non-trivial constraint on model building in this case.
\end{abstract}


\maketitle
\section{Introduction}

Microscopic nature of dark matter is one of the central open questions in fundamental physics which cannot be addressed within the Standard Model (SM). Many theoretical ideas have been suggested, and an extensive experimental effort is under way to test some of the proposals~\cite{Battaglieri:2017aum}. While the precise nature of the dark matter sector varies greatly among the proposed models, all of them postulate that dark matter consists of point-like particles (e.g. WIMPs or axions), their bound states (e.g. dark atoms), or particle-like extended objects (monopoles, Q-balls, etc.), both today and throughout its cosmological history. However, viable extensions of the SM exist in which new physics sectors do not contain spatially localized particle-like excitations at all~\cite{Georgi:2007ek}. A well-known example is a conformal field theory (CFT)~\cite{CFT:BigBook, Ginsparg:1988ui, Rychkov:2016iqz}, where scale invariance precludes the existence of stable finite-size states. In this letter, we show how dark matter can arise from a new physics sector which is described by a CFT throughout most of its cosmological history.                      

An immediate objection to the idea of dark matter made out of CFT ``stuff" is that conformal invariance dictates that the energy density of such stuff redshifts like radiation ($\rho \propto a^{-4}$), rather than non-relativistic matter ($\rho \propto a^{-3}$), as the universe expands. However, in any phenomenologically viable model, conformal invariance is at most approximate and must be broken to some degree. In particular, any interactions of the CFT sector with the non-conformally-invariant SM inevitably break the symmetry. Generically, such effects induce a ``gap" mass scale, below which the sector is no longer conformal and its spectrum consists of spatially localized particle degrees of freedom. Below, we will discuss a scenario in which dark matter production in the early universe occurs at temperatures above the gap scale, so that throughout the production process the dark sector can be well approximated by a CFT. At the same time, the gap scale, which is induced by cosmological phase transitions in SM, is sufficiently large so that the dark sector behaves as non-relativistic matter during CMB decoupling, structure formation, and today, as required by observations.        

\section{Particle Physics Framework}

We extend the SM by postulating a dark sector, whose fields do not carry SM gauge charges. The dark sector is assumed to be invariant under the  conformal group. It is coupled to the SM via 
\beq
{\cal L}_{\rm int} = \frac{\lambda_{\scriptstyle_{\rm CFT}}}{\Lambda_{\scriptstyle_{\rm CFT}}^D} \,\Os \Oc\,,
\eeq{interaction} 
where $\Os$ is a gauge-invariant operator consisting only of SM fields, $\Oc$ is an operator within the dark-sector CFT, and $\Lambda_{\scriptstyle_{\rm CFT}}$ is the energy scale where the CFT is replaced by its ultraviolet (UV) completion. We will consider the regime of small Wilson coefficient $\lambda_{\scriptstyle_{\rm CFT}}\ll 1$, where the conformal symmetry breaking introduced by Eq.~\leqn{interaction} can be treated as a (technically natural) small perturbation. A UV completion of the CFT that naturally generates $\lambda_{\scriptstyle_{\rm CFT}}\ll 1$ is discussed below. If $\Os$ and $\Oc$ have scaling dimensions $\ds$ and $d$, respectively, then
\beq
D = \ds+d-4.
\eeq{dims}  
The dark-sector CFT may be strongly coupled, resulting in large anomalous dimensions and non-integer $d$. A simple and predictive scenario for CFT breaking in the infrared (IR) is to consider SM operators with $\left< \Os \right> \not= 0$, which automatically triggers such breaking through the interaction term in Eq.~\leqn{interaction} if $\Oc$ is relevant, $d<4$. In this scenario $\Oc$ must be a scalar operator, and CFT unitarity then requires $d\geq 1$. Two obvious choices for $\Os$, which will be our focus in this work, are:

\begin{itemize}
	
	\item Higgs portal: $\Os=H^\dagger H$ ~~~~~~~~($\ds=2$);
	
	\item Quark portal: $\Os=H Q_L^\dagger q_R$ ~~~~($\ds=4$).
	
\end{itemize}
For both portals, $\left< \Os\right> \not= 0$ in the infrared (IR), due to the Higgs vacuum expectation value (vev) and the QCD chiral condensate. Conformal symmetry is broken at a scale $\gap$, for which dimensional analysis gives 
\beqa
\gap &\sim& \left(\frac{\lambda_{\scriptstyle_{\rm CFT}} \,v^2}{\Lambda_{\scriptstyle_{\rm CFT}}^{d-2}}\right)^{\frac{1}{4-d}}~~~~~~~~~~~~({\rm Higgs~portal}); \CR
\gap &\sim& \left(\frac{\lambda_{\scriptstyle_{\rm CFT}} \,v \Lambda_{\scriptstyle_{\rm QCD}}^3 }{\Lambda_{\scriptstyle_{\rm CFT}}^d}\right)^{\frac{1}{4-d}}\,~~~~~({\rm quark~portal}).
\eeqa{gapHH}
Once the conformal symmetry is broken, the spectrum consists of particle-like excitations with masses $\sim\gap$ which can be thought of as bound states of the original CFT degrees of freedom. We assume that one of these excitations is stable on cosmological time scales, for example, due to a discrete symmetry. This is the particle that will play the role of dark matter (DM). Regarding the DM particle mass, we will consider two possibilities. One is that the DM particle is a generic bound state, with mass $\mdm = \gap$ (up to order-one factors). The second one is that the DM particle is a pseudo-Goldstone boson (PGB) of an approximate global symmetry spontaneously broken at $\gap$, similar to pions in QCD. In this case, $m_{\rm DM} \ll \gap$ is natural, with the DM mass dictated by the amount of explicit symmetry breaking.       

The strongly-coupled, conformally invariant dark sector of our model can arise from a weakly-coupled, asymptotically free theory in the UV. This can be a simple $SU(N_c)$ gauge theory with $N_F$ fermion flavors ${\cal Q}_i$, which flows towards a Banks-Zaks (BZ) fixed point in the infrared~\cite{Banks:1981nn}. The theory becomes strongly coupled, and approximately conformal, at a scale $\Lambda_{\rm CFT}$. The interaction with the SM starts out as
\beq
{\cal L}_{\rm int} = \frac{1}{M_U^{d_U}} \mathcal{O}_{\rm SM} \mathcal{O}_{\rm BZ}\,,
\eeq{int_BZ} 
where $\mathcal{O}_{\rm BZ}$ is a gauge-invariant operator in the dark-sector gauge theory, with a scaling dimension $d_{\rm BZ}$. Since $d_U=d_{\rm SM}+d_{\rm BZ}-4>0$, this interaction is irrelevant, and $M_U \gg \Lambda_{\scriptstyle_{\rm CFT}}$ is required for consistency. At the scale $\Lambda_{\rm CFT}$, this interaction term is matched to the one in Eq.~\leqn{interaction}, with 
\beq
\lambda_{\scriptstyle_{\rm CFT}} \sim \left( \frac{\Lambda_{\scriptstyle_{\rm CFT}}}{M_U}\right)^{d_U}  \ll 1.
\eeq{scale_relation}  
For example, we may consider the quark bilinear operator $\mathcal{O}_{\rm BZ}={\cal Q}_i^\dagger {\cal Q}_i$, with $d_{\rm BZ}=3$.   
Our scenario requires that this operator acquire a large anomalous dimension, $\gamma_{\rm BZ} \sim {\cal O}(1)$, at the IR fixed point. Such large anomalous dimensions, with the sign consistent with our scenario, have been observed in lattice studies of $SU(3)$ gauge theory with $N_f=10$~\cite{Appelquist:2012nz} and $N_f=8$~\cite{Aoki:2013xza,Aoki:2014oha,Appelquist:2014zsa,Appelquist:2016viq,Appelquist:2017wcg}, as well as in analytic scheme-independent calculations at higher orders in perturbation theory~\cite{Ryttov:2017kmx,Ryttov:2017dhd}.

We note that in the case of Higgs portal for the special value $d=2$, the theory we consider bears some superficial resemblance to the ``scale-invariant" dark sectors consisting of massless, weakly-coupled fields, considered in previous studies (see for example~\cite{Kang:2014cia, Heikinheimo:2017ofk}). However, the physics is completely different: our dark sector is strongly coupled, the operator $\Oc$ has a non-perturbatively large anomalous dimension (which just happens to be an integer at this special point), and is genuinely (not just classically) conformally invariant.     

Since $\Oc$ is a relevant operator, an additional symmetry must be invoked to avoid its appearance in the CFT Lagrangian, which would lead to incalculable IR breaking of the CFT that would generically dominate the effect of Eq.~\leqn{interaction}. This may for example be a ${\cal Z}_2$ symmetry under which $\Oc$ is odd, explicitly broken only by the interaction with the SM.\footnote{Even with discrete symmetry, loops involving SM particles will induce IR breaking of the CFT. For example, in the case of Higgs portal, $\delta {\cal L}_{\scriptstyle_{\rm CFT}} \sim \frac{\lambda_{\scriptstyle_{\rm CFT}}}{\Lambda_{\scriptstyle_{\rm CFT}}^D} \, \frac{\Lambda_{\scriptstyle_{\rm SM}}^2}{16\pi^2} \Oc$, where $\Lambda_{\scriptstyle_{\rm SM}}$ is the scale where quadratic divergence in the Higgs loop is cut off, for example by compositeness or supersymmetry. This effect is subdominant to the breaking due to Higgs vev as long as $\Lambda_{\scriptstyle_{\rm SM}} \lesssim 4\pi v$, the usual condition for naturalness of the weak scale.} However, the operator product expansion (OPE) of $\Oc \times \Oc$ generally contains singlet scalar operators, which are even under ${\cal Z}_2$~\cite{Poland:2018epd}. Numerical CFT bootstrap provides an upper bound on the dimension of the lowest singlet scalar operator in the OPE~\cite{Poland:2011ey, Poland:2018epd}. Requiring that no 
${\cal Z}_2$-even relevant operators are generated implies $d > 1.61$~\cite{Poland:2011ey}. Note that this bound is model-dependent and may be avoided if a larger symmetry is used, or if operator coefficients are fine-tuned.

Note that with the exception of the IR breaking of conformal symmetry, the field theory model considered here is identical to Georgi's ``unparticle" framework \cite{Georgi:2007ek, Grinstein:2008qk}), with SM coupling to a scalar CFT operator. (For earlier works on cosmology with unparticles, see~\cite{Davoudiasl:2007jr, Chen:2007qc, Kikuchi:2007az, McDonald:2008uh, Grzadkowski:2008xi}.) In other words, the dark sector behaves as an unparticle at energy scales above $\gap$ and below $\Lambda_{\rm CFT}$.      

\section{Cosmological Evolution}

We assume that after the end of inflation, the inflaton decay reheats the SM sector to a temperature $T_R$, but the dark sector is not reheated due to absence of a direct coupling to the inflaton. As the universe expands and cools after reheating, collisions and decays of SM particles gradually populate the dark sector. Assuming $\gap \ll T_R < \Lambda_{\rm CFT}$, this process occurs via production of CFT stuff (``unparticles"). The dark sector cannot be described by Boltzmann equations, since the concept of particle number density is not applicable in the CFT. However, since the dark sector has many degrees of freedom and they interact strongly among themselves, it will be in a spatially isotropic thermal state. Rotational symmetry dictates that the energy-momentum tensor of this state has the form $T^{\mu\nu}= {\rm diag}(\rhoc,-\Pc,-\Pc,-\Pc)$, while conformal invariance further requires $\Pc=\frac{1}{3}\rhoc$. The CFT energy density is given by 
\beq
\rhoc = A \Tc^4,
\eeq{Tcrhoc}
where $\Tc$ is the temperature of the CFT sector, and $A$ is an order-one model-dependent constant. We will study a scenario where $\Tc\ll T$ at all times, where $T$ is the SM plasma temperature; at the same time, $\Tc> \gap$ during the period when the dark sector is populated, so that the CFT description is appropriate.   

On the SM side, the particle number is well-defined and the Boltzmann equations have the usual form, with collision terms describing the loss of SM particles due to annihilations (SM+SM$\to$ CFT) and decays (SM$\to$ CFT), and their creation due to inverse processes. The evolution of the SM energy density $\rho_{\rm SM}$ follows from the Boltzmann equations: 
\beqa
& & \hskip-1cm \frac{d\rhos}{dt} + 3H (\rhos + \Ps) =  \CR & & -\Gamma_E({\rm SM}\rightarrow {\rm CFT}) + \Gamma_E({\rm CFT}\rightarrow {\rm SM})\,,
\eeqa{BeqSM}
where $H$ is the Hubble expansion rate, and $\Gamma_E$ are energy transfer rates per unit volume. In our scenario, the CFT sector will always remain at densities far below equilibrium with the SM, and $\Gamma_E({\rm CFT}\rightarrow {\rm SM})$ can be safely neglected. The energy transfer rate from SM to CFT is given by
\beqa
\Gamma_E({\rm SM}\rightarrow {\rm CFT}) &=& \sum_{i,j} n_i n_j \langle \sigma(i+j\to {\rm CFT}) v_{\rm rel} E\rangle \CR& & + \sum_i n_i \langle \Gamma(i\to {\rm CFT}) E\rangle,
\eeqa{GammaE}
where the sums run over all SM degrees of freedom coupled to the CFT. The cross sections and decay rates can be evaluated using the technique of Georgi~\cite{Georgi:2007ek, Grinstein:2008qk}. For example, with the Higgs portal, the Higgs decay contribution is given by 
\beq
n_h \langle \Gamma(h\to {\rm CFT}) E\rangle \,=\, \frac{f_d \lambda_{\scriptstyle_{\rm CFT}}^2 v^2 m_h^{2(d-1)} T}{\Lambda_{\scriptstyle_{\rm CFT}}^{2d-4}}\,K_2(m_h/T)\,
\eeq{hh_decay}
where $m_h$ is the Higgs boson mass, $f_d=2^{-2d}\pi^{1/2-2d}\Gamma(d+1/2)/(\Gamma(d-1)\Gamma(2d))$, and $K_2(x)$ is the modified Bessel function of the second kind. The annihilation contribution (when $T\gg m_h$) is given by
\beq
n_h^2 \langle \sigma(h h \rightarrow \mathrm{CFT})\; v_{\rm rel} \; E  \rangle = \lambda_{\scriptstyle_{\rm CFT}}^2 \, \frac{d}{2(2\pi)^{2d+1}}\;  \frac{T^{2d+1}}{\Lambda_{\scriptstyle_{\rm CFT}}^{2d-4}}\,.
\eeq{hh_scat_gamma}  
The CFT sector is populated at the time when the energy density is dominated by relativistic SM matter, $\Ps=\frac{1}{3}\rhos$, so that SM and CFT energy densities redshift in the same way. The total energy of the two sectors can only change due to work done against the expansion of the universe:
\beq
\frac{d}{dt}\left(\rhoc+\rhos\right) + 4H \left(\rhoc+\rhos\right)  = 0.
\eeq{total}
Subtracting Eq.~\leqn{BeqSM}, we find that the CFT energy density evolves according to 
\beq
\frac{d\rhoc}{dt} + 4H \rhoc  =  \Gamma_E({\rm SM}\rightarrow {\rm CFT})\,,
\eeq{BeqCFT}
with the initial condition $\rhoc=0$ at $T=T_R$. 

\begin{figure}[t!]
	\begin{center}
		\includegraphics[width=8.6cm]{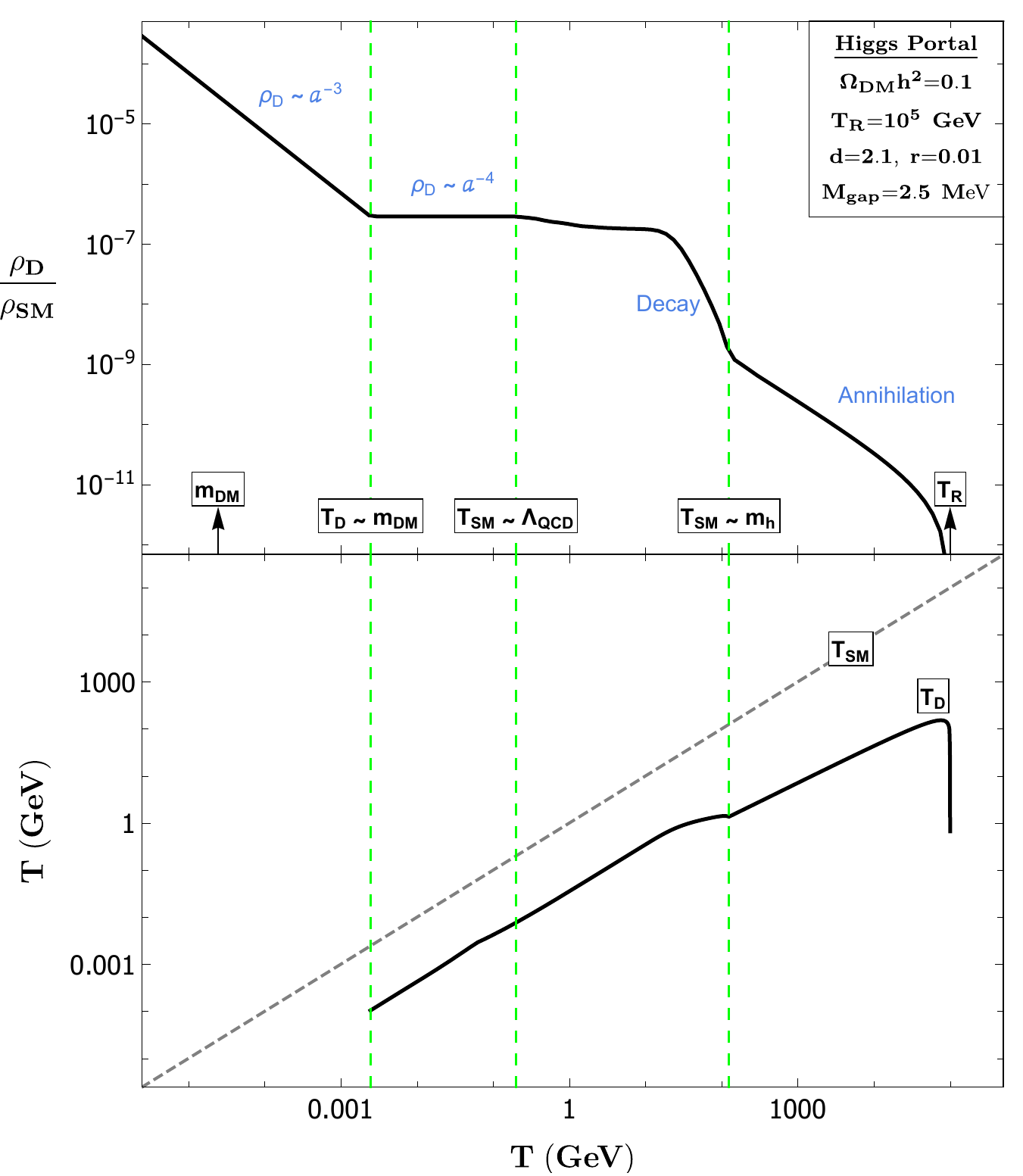}
		\vspace{-2mm}
		\caption{Top panel: Energy density in the CFT plasma or dark matter particles, as a function of the SM plasma temperature $T$, in the Higgs portal scenario with $\Lambda_{\scriptstyle_{\rm CFT}}=1.2\times 10^8$~GeV and $M_U=10^{17}$~GeV. Bottom panel: Evolution of the CFT plasma temperature, as a function of $T$, for the same parameters. At all times, $\Tc\ll T$, as required for the self-consistency of our calculations.}
		\label{fig:yield}
	\end{center}
\end{figure}

With minor simplifying assumptions, such as ignoring the masses of colliding SM particles and the temperature dependence of the effective number of SM degrees of freedom $g_*$, Eq.~\leqn{BeqCFT} can be solved analytically (see Supplemental Material for details). The qualitative behavior of $\rhoc$ with temperature is dictated by the dimension $d$ of the operator $\Oc$. For $d$ above the critical dimension $d_*$, most of the CFT energy is produced at high temperatures, close to $T_R$, by pair-annihilations of SM particles; the decay contribution, if present, is subdominant.\footnote{For earlier work on freeze-in in the UV sensitive regime, see for e.g.~\cite{Elahi:2014fsa}.} On the other hand, for $d<d_*$, contributions from both pair-annihilations and decays (if present) grow with decreasing $T$. The resulting CFT density is IR dominated and can be calculated without knowledge of UV quantities such as $T_R$, as in the freeze-in scenario of Ref.~\cite{McDonald:2001vt, Hall:2009bx} (for review of variations of freeze-in models and phenomenology, see~\cite{Bernal:2017kxu}). We will focus on this case for the rest of the paper. We call this scenario {\it Conformal Freeze-In}, or COFI. The critical dimensions are $d_*=5/2$ for the Higgs portal, and $d_*= 3/2$ for the quark portal. 

For both portals, production of CFT energy effectively ceases soon after the SM temperature drops below the mass of the particle coupled to the CFT: $T\sim m_h$ for the Higgs portal\footnote{In the Higgs portal case, there is a residual quark/gluon-CFT interaction below the weak scale, induced through integrating out the Higgs. Numerically, its effect is subdominant.} and $T\sim \Lambda_{\scriptstyle_{\rm QCD}}$ for the quark portal. After that, $\rhoc$ redshifts as $a^{-4}$, until the CFT temperature drops to $\Tc \sim \gap$. At that time, conformal symmetry is broken and a confining phase transition takes place. We assume that all of the energy stored in the CFT sector is transferred to DM particles, on a time scale short compared to Hubble at that time. If $\mdm\ll\gap$, the dark matter energy density will continue to redshift as radiation until its temperature drops below $\mdm$, after which it behaves as non-relativistic matter. With these assumptions, the relic density can be estimated analytically (see Supplemental Material). For example, for the Higgs portal, the relic density is dominated by the Higgs decay contribution, and is given by 
\beq
\frac{\Omega_{\scriptstyle_{\rm DM}} h^2}{0.1} = \left[ \frac{\mdm}{1\mev} \right] \left[ \frac{\left(A \, f_d^3 \, g_*^{-9/2}\right)^{1/4}}{ 10^{-5}} \right] 
\left[ \frac{\left( \frac{\gap}{m_h} \right)^{(6-\frac{3d}{2})}}{10^{-12}} \right],\,
\eeq{anal_HP}
where $g_*\sim 100$ is the number of relativistic SM degrees of freedom at the weak scale. An example of the evolution of CFT/DM energy density, for the Higgs portal scenario and parameters that provide the observed DM relic density, is shown in Fig.~\ref{fig:yield}. This and all figures below are based on full numerical solutions of Eq.~\leqn{BeqCFT}, which is in good agreement with Eq.~\leqn{anal_HP}. 
 
\section{Dark Matter Phenomenology}

\begin{figure}[t!]
	\begin{center}
		\includegraphics[width=8.5cm]{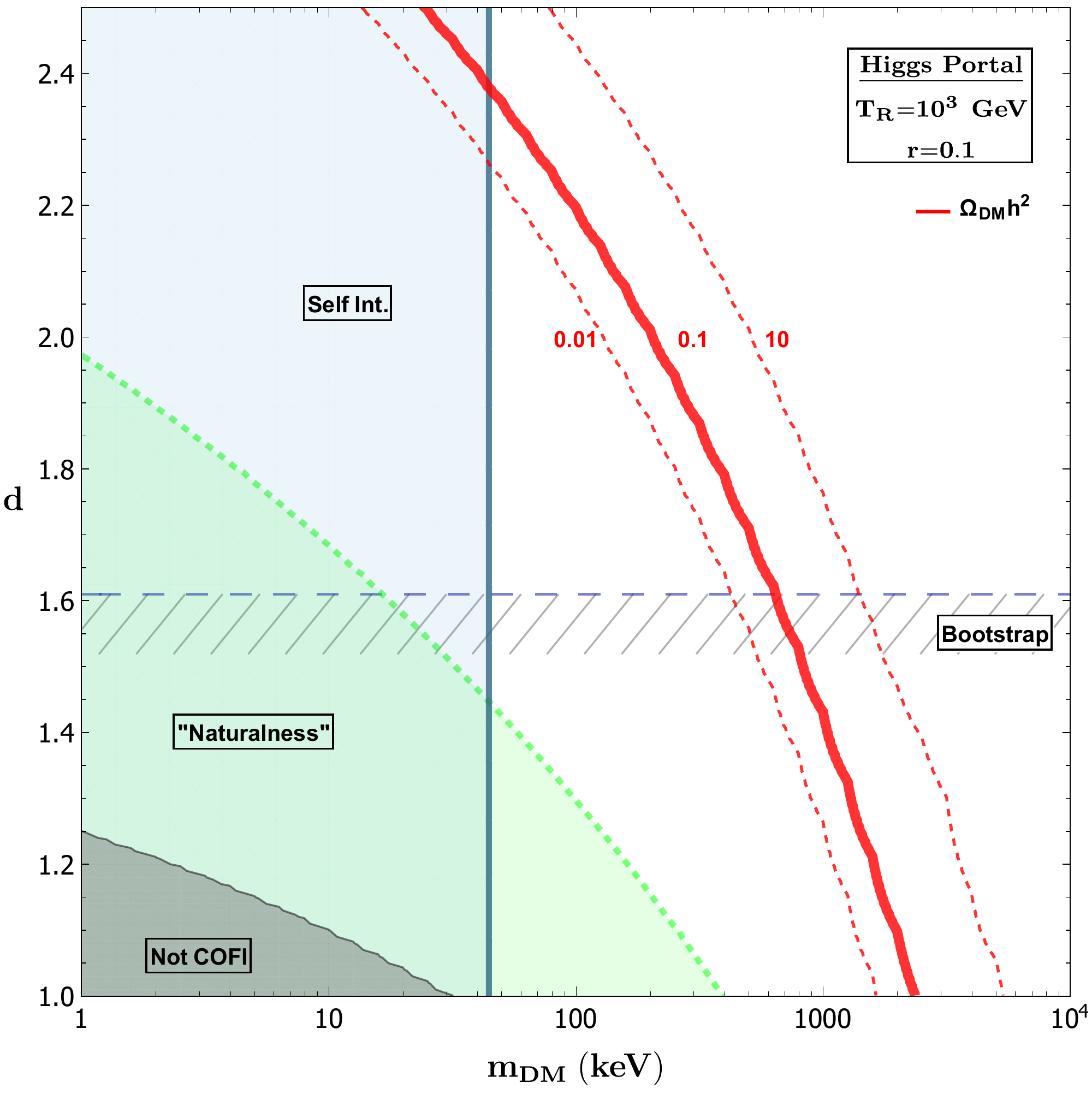}
		\vspace{-2mm}
		\caption{Dark matter relic density contours (red) and observational/theoretical constraints, in the Higgs portal model. Thick red line indicates parameters where the observed dark matter abundance is reproduced.}
		\label{fig:DMhiggs}
	\end{center}
\end{figure}

Since the dark matter relic density is determined by IR physics, and the dark matter mass and its interaction strength are related through Eqs.~\leqn{gapHH}, the COFI scenario is remarkably predictive. In particular, there is a nearly universal relationship between the dark matter relic density and the gap scale, with only a mild dependence on other parameters. In the case of the Higgs portal, the observed relic density is reproduced for $\gap\sim 1-10$~MeV, while for the quark portal, it is $\gap\sim 10-100$~keV.

\begin{figure}[t!]
	\begin{center}
		\includegraphics[width=8.5cm]{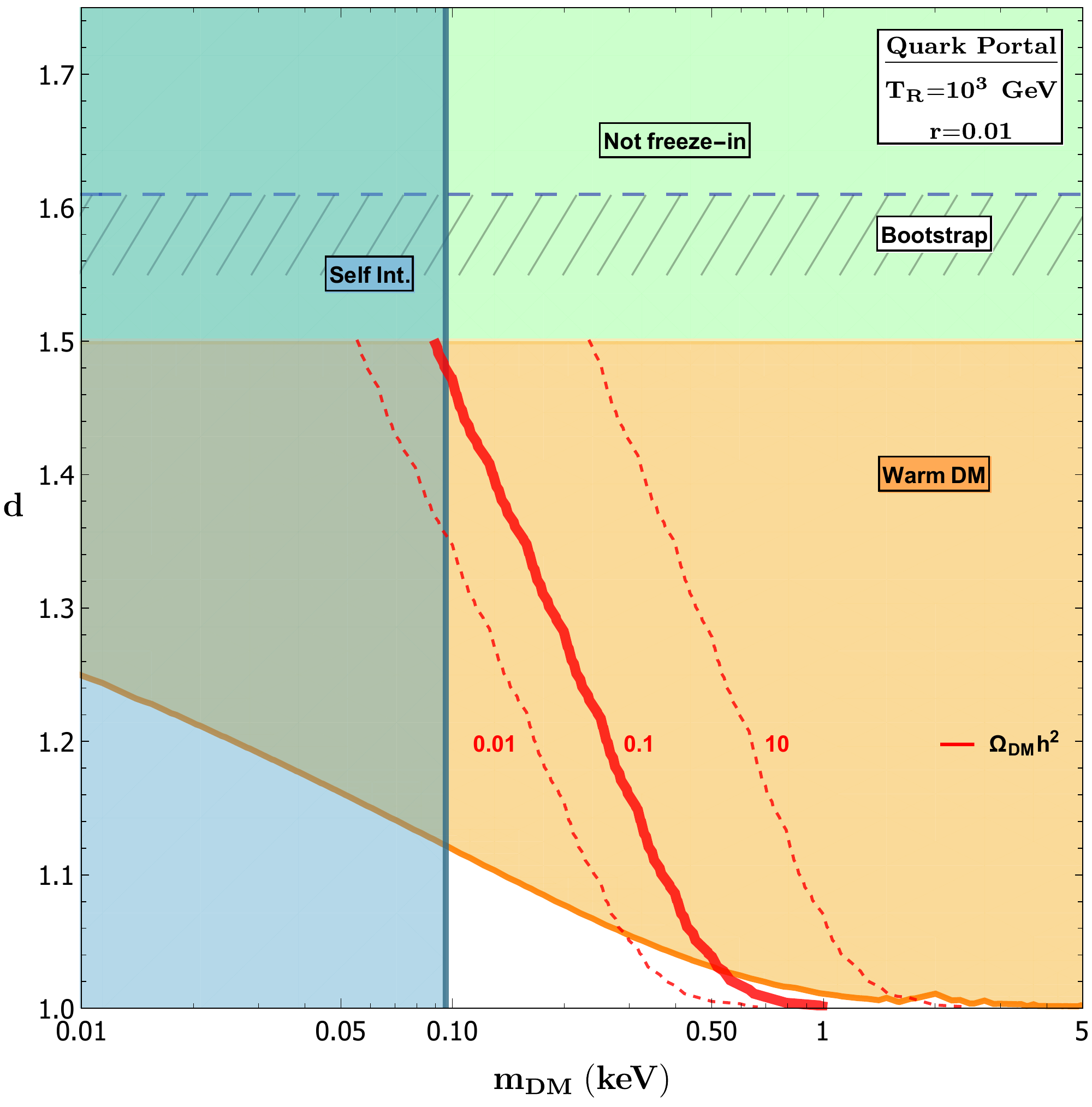}
		\vspace{-2mm}
		\caption{Dark matter relic density contours (red) and observational/theoretical constraints, in the quark portal model. Thick red line indicates parameters where the observed dark matter abundance is reproduced.}
		\label{fig:DMquark}
	\end{center}
\end{figure}

If the DM particle is a generic bound state of a strongly-coupled theory with $\mdm\sim\gap$, its elastic self-scattering interaction cross section can be estimated as $\sigma_{\scriptstyle_{\rm self}} \approx 1 / \left( 8\pi \gap^{2} \right)$. This is far too large, in both Higgs and quark portal scenarios, to be consistent with bounds from galaxy clusters such as the Bullet cluster~\cite{Markevitch:2003at, Lin:2019uvt}. We therefore consider the case where the DM is a PGB, with mass ratio $r=\mdm/\gap \ll 1$. The self-scattering cross section scales as
\beq
\sigma_{\scriptstyle_{\rm self}} \sim \frac{\mdm^6}{8\pi \gap^8} = \frac{r^6}{8\pi \gap^{2}} ,
\eeq{self}
where we assumed that self-scattering is mediated by states with masses $\sim \gap$ (e.g. the counterparts of the $\rho$ meson of QCD), and derivative couplings of the PGB have been taken into account. Modest values of $r\sim 0.01-0.1$ are sufficient to avoid the self-interaction bounds for parameters with viable $\Omega_{\scriptstyle_{\rm DM}}$. This is illustrated in Figs.~\ref{fig:DMhiggs} and~\ref{fig:DMquark}.  

Another important phenomenological constraint is that the DM should be cold, {\it i.e.} remain non-relativistic during structure formation \cite{Irsic:2017ixq}. This constraint is somewhat weaker than in the case of SM sterile neutrinos, $m\gsim 5$ keV, because the CFT sector is colder than the SM. Nevertheless, for the quark portal, the warm DM bound rules out a significant part of the parameter space; see Fig.~\ref{fig:DMquark}. For the Higgs portal, the DM is heavier and this bound is irrelevant.                

There are many experimental and observational constraints on the strength of CFT/DM coupling to the SM. These include LHC searches for unparticles produced in $q\bar{q}$ annihilations~\cite{Sirunyan:2017onm, Khachatryan:2014rra} or Higgs decays~\cite{Khachatryan:2016whc, Sirunyan:2018owy}; bounds on invisible meson decays involving unparticles from the B-factories~\cite{Lees:2013kla, Artamonov:2008qb, Krnjaic:2015mbs}; supernova SN1987a energy loss and stellar cooling due to unparticle or DM emissions~\cite{Turner:1987by, Davoudiasl:2007jr, Fradette:2018hhl}; 
modification of the ionization history due to energy injection by late-time DM annihilations~\cite{Chen:2003gz, Pierpaoli:2003rz, Slatyer:2012yq, Slatyer:2015jla}; diffuse $X$- and Gamma-ray backgrounds~\cite{Essig:2013goa}; and spectral distortion of the CMB blackbody distribution by early energy injection~\cite{Chluba:2011hw}. We checked that our scenario is easily consistent with all these bounds, due to a highly suppressed coupling between the dark sector and the SM. 
Note also that the Big Bang Nucleosynthesis (BBN) bound on the number of new light degrees of freedom~\cite{Cyburt:2015mya} does not apply, because $\Tc\ll T$ at the time of BBN.   

On the theoretical side, there are several consistency conditions that may further constrain the parameter space, as shown in Figs.~\ref{fig:DMhiggs} and~\ref{fig:DMquark}. The ``naturalness" bound stems from requiring that if $T_R<\Lambda_{\scriptstyle_{\rm CFT}}$, then $M_U< M_{\rm Pl}$. ``Bootstrap" condition, $d>1.61$,  
is explained in the paragraph below Eq.~\leqn{scale_relation}. Both these bounds are model-dependent, and may be modified or eliminated by the choice of a UV completion of the CFT and symmetry charge assignment for $\Oc$, respectively.

We conclude that for the Higgs portal, the conformal freeze-in (COFI) dark matter scenario is easily viable, consistent with all observational and theoretical constraints. This scenario contains a novel DM candidate, a CFT bound state, whose mass is predicted to be in the experimentally interesting sub-MeV range. The lower bound on the DM mass is about a keV, where warm dark matter constraint becomes insurmountable. 
For the quark portal, the situation is more constrained. A small sliver of parameter space, with $d$ close to 1, is observationally viable, but inconsistent with the bootstrap condition in its simplest form. It is interesting that viability of the DM candidate in this case hinges upon highly non-trivial intrinsic properties of the CFT.     

~\\
{\em Acknowledgments ---} We are grateful to Damon Binder, Michael Geller, Thomas Hartman, Luca Iliesiu, Hyung Do Kim, Eric Kuflik, Ofri Telem, and Eliott Rosenberg for useful discussions. We acknowledge discussions with participants of ``The 4th NPKI workshop", and ``Searching for new physics - Leaving no stone unturned!" at the University of Utah. SH acknowledges hospitality and financial assistance by the Munich Institute for Astro- and Particle Physics (MIAPP) of the DFG Excellence Cluster Origins (www.origins-cluster.de). This work is supported by the U.S. National Science Foundation through grant PHY-1719877, and by Cornell University through the Hans Bethe Postdoctoral Fellowship (SH) and Cornell Graduate Fellowship (GK). 

~\\


\newpage

\section{Critical Dimension ($d^*$)}

Here, we derive the critical dimension ($d^*$) of the CFT operator, below which dark sector energy production is IR-dominated and independent of early universe parameters such as the reheating temperature.

In the examples considered in this paper, $\SM\,+\,\SM\,\rightarrow\,\mathrm {CFT}$ and $\SM\,\rightarrow\,\CFT$ processes are the only relevant modes of production. Others, such as $3\, \SM\,\rightarrow\,\mathrm {CFT}$ processes, are disfavored strongly by phase space suppression and can be ignored. In the case of Standard Model decays, production is always UV-insensitive, as the bulk of it occurs when the temperature is of the same order as the mass of the SM particle. On the other hand, production from $2\,\rightarrow\,\mathrm {CFT}$ processes may occur primarily either in the UV or in the IR, depending on the dimension of the CFT operator coupled to the Standard Model. 

Consider the following coupling, as shown in Eq.~\leqn{interaction}:
\beq
{\cal L}_{\rm int} = \frac{\lambda_{\scriptstyle_{\rm CFT}}}{\Lambda_{\scriptstyle_{\rm CFT}}^{(D-4)}} \,\Os \Oc\,,
\eeq{int} 
where $D= d+\ds$ as before.

We will calculate the critical dimension for the case of $2 \rightarrow \CFT$ processes as follows. From dimensional analysis, it is easy to relate the collisional term in the Boltzmann equation for two SM particles to the SM temperature and the CFT scale, since the cross-section under consideration will have a factor of $1/\L^{2(D-4)}$. Without any dimensionless factors, we get
\beqa
& n_\mathrm{SM}^2 \sim T^6 \; , \ \langle \, \sigma \, v \, E \, \rangle \sim \frac{T^{2D-9}}{\L^{2(D-4)}} \nonumber\\
& \Rightarrow \Gamma_E (\SM\,\rightarrow\,\CFT) \sim \frac{T^{2D-3}}{\L^{2D-8}}.
\eeqa{NDA1}
The dark sector energy density can be estimated by integrating the Boltzmann equation (Eq.~\leqn{BeqCFT}) with the added constraint that there is no dark sector energy above the reheating temperature. Thus, for temperatures below reheating ($T<T_R$), 
\beqa
\rho_\mathrm{CFT} \sim T^4 \,\left(\frac{T^{-\eta} - T_R^{-\eta}}{\eta}\right)\ \mathrm{with\ } \eta = 9 - 2(d+\ds).\ \ 
\eeqa{NDA2}

For $\eta > 0 \Rightarrow d < d^* = 9/2 - d_\SM$, the reheating temperature term will be negligible due to its negative exponent, and most of the dark sector energy density will be produced at lower temperatures. Thus, it will be IR-dominated, and relic density will not depend strongly on UV parameters such as the reheating temperature. On the other hand, for $d > d^*$, the reheating temperature plays a relevant role: energy density production peaks at $T_R$, and then dilutes due to the expansion of the universe.

In the Higgs portal case, we checked that, as long as $d<d^*$, Higgs decay is the dominant process in production of CFT energy density. Above the critical dimension however, this is not necessarily true, as the scattering contribution may dominate for a sufficiently high reheating temperature.

\section{Derivation of Relic Density in the Higgs Portal}

In this section, we show a brief derivation of Eq.~\leqn{anal_HP}, that relates observed dark matter relic density to parameters in the theory. In addition, the computation for Eq.~\leqn{hh_decay} is shown in more detail.

In the Higgs portal case, as mentioned before, below the critical dimension $d*\,=\, 5/2$, dark matter production is dominated by the Higgs decay process. At temperatures below the electroweak phase transition, the effective interaction between the dark sector and the SM becomes, 
\beq
{\cal L}_\mathrm{int} = \frac{\l}{\L^D} \frac{v}{\sqrt{2}}\ h\ \Oc.
\eeq{L_int_decay}
The energy transfer rate through this process is given by Eq.~\leqn{hh_decay} and can be computed as follows:
\beqa
& &n_h \langle \; \Gamma(h \rightarrow \CFT) \; E  \rangle \nonumber\\
& = &\iint \d\Pi_h \d\Pi_\CFT f_h (2\pi)^4 \delta^4(p_h - P) E_h \vert \mathcal{M} \vert^2. \ \ 
\eeqa{hh_1}
Here and below, $P = p_\CFT$ is the momentum carried by the dark sector. The phase space for the CFT sector is chosen to be identical to that of ``unparticles" as prescribed by Georgi in~\cite{Georgi:2007ek}. Using Georgi's notation, we have,
\beqa
&& n_h \langle \; \Gamma(h \rightarrow \CFT) \; E  \rangle \nonumber\\
&& = \iint \frac{\d^3 \vec{p}_h}{(2\pi)^3 2E_h}\ \frac{\d^4 P}{(2\pi)^4} \mathrm{e}^{-\beta E_h} (2\pi)^4\ \delta^4(p_h - P) \nonumber\\
&& \hspace{8em}  \times \ A_d \ (P^2)^{d-2} \ E_h\ \frac{v^2}{4}\ \frac{\l^2}{\L^{2d-4}} \nonumber\\ 
&&= \frac{A_d \; v^2 \; \l^2}{4\L^{2d-4}} (m_h^2)^{d-2} \int \frac{\d^3 \vec{p}_h}{2(2\pi)^3} \,\exp(\sqrt{\vert\vec{p}_h\vert^2 + m_h^2}), \nonumber\\
&&\hspace{12em} 
\eeqa{hh_2}
where
\beq
A_d = \frac{16\pi^{5/2}}{(2\pi)^{2d}}\frac{\Gamma(d+1/2)}{\Gamma(d-1)\Gamma(2d)}.
\eeq{}
Setting $p=\vert\vec{p}_h\vert$ and simplifying gives
\beqa
&&n_h \langle \; \Gamma(h \rightarrow \CFT) \; E  \rangle = \frac{A_d \; v^2 \; \l^2 \; (m_h^2)^{d-2}}{4\L^{2d-4}} \nonumber\\
&& \hspace{5em} \times \int 4 \pi p^2 \frac{\d p}{2(2\pi)^3}\, \exp(-\beta \sqrt{p^2 + m_h^2} ) \nonumber\\
&& = \frac{A_d \; v^2 \; \l^2 \; (m_h^2)^{d-2}}{32 \pi^2 \L^{2d-4}}  \int p^2 \,\d p\, \,\exp(-\beta \sqrt{p^2 + m_h^2} ). \nonumber\\
&&\hspace{12em} 
\eeqa{hh_3}
The integral represents a Bessel function of the second kind. Additionally, in our notation, $f_d = A_d/16\pi^2$. Thus, on simplifying, we get,
\beq
n_h \langle \Gamma(h\to {\rm CFT}) E\rangle \,=\, \frac{f_d \lambda_{\scriptstyle_{\rm CFT}}^2 v^2 m_h^{2(d-1)} T}{\Lambda_{\scriptstyle_{\rm CFT}}^{2(d-4)}}\,K_2(m_h/T).
\eeq{hh_decay2}
The CFT energy density at any point in time (as a function of the Standard Model bath temperature) can be obtained by integrating the Boltzmann equation given in Eq.~\leqn{BeqCFT}. To get a simple estimate, it suffices to do this calculation in the relativistic approximation where the Higgs is assumed to be massless and is described by a Maxwell-Boltzmann distribution. The process roughly starts around the electroweak scale $\sim v$ and continues till the SM temperature reaches the Higgs mass. 

In the relativistic approximation (i.e., taking the limit $m_h \rightarrow 0$ in the thermal average calculation), the energy transfer rate in this process is given by,
\beq
n_h \langle \Gamma(h\to {\rm CFT}) E\rangle \,=\, 2 f_d \, \l^2 \, v^2 \, \frac{m_h^{2d-4}}{\Lambda_{\scriptstyle_{\rm CFT}}^{2d-4}}\,T^3.
\eeq{hh_decay_rel}
We integrate the Boltzmann equation with this collisional term, ignoring the temperature dependence of $g_*$ for now, and enforcing the condition that decays are inactive above the electroweak scale. Thus, we have,
\beq
\rhoc (T) \,=\, \frac{2 M_* f_d \lambda_{\scriptstyle_{\rm CFT}}^2 }{3 \sqrt{g_*(T)} v}\, \left( \frac{m_h}{\Lambda_{\scriptstyle_{\rm CFT}}} \right)^{2d-4} \,T^4\left( \frac{v^3}{T^3}-1 \right),
\eeq{rho_CFT_decay}
where $M_* = 3\sqrt{5}/(2\pi^{3/2}) \, M_{pl}$, comes from the definition of Hubble as $H = \sqrt{g_*}\; T^2/M_*$.

At $T \sim m_h$, as the Higgs falls out of the thermal bath, this process becomes exponentially suppressed, and further production of dark sector energy can be neglected for this analysis. The energy density present in the dark sector then redshifts like radiation ($\rho \propto a^{-4}$) until its temperature $\Tc$ becomes comparable to the mass of the dark matter candidate. After this point, it redshifts like matter ($\rho \propto a^{-3}$) as required.

Thus,
\beq
\rhoc (m_h) \,  =\, \frac{2 M_* f_d \l^2 }{3 \sqrt{g_*(m_h)} v}\, \frac{m_h^{2d}}{\L^{2d-4}}\, \left( \frac{v^3}{m_h^3}-1 \right), 
\eeq{}
and
\beqa
&\rhoc (T_m) \,  =\, \frac{2 M_* f_d \l^2 \, g_*(T_m)}{3 (g_*(m_h))^{3/2} v }\,\left( \frac{m_h}{\L} \right)^{2d-4} \nonumber \\ 
& \hspace{10em} \times \left( \frac{v^3}{m_h^3}-1 \right) T_m^4,
\eeqa{rho_CFT_end}
where $T_m$ is the \textit{SM temperature} at which the \textit{dark sector temperature} drops to the mass of the dark matter candidate. From Eq.~\leqn{Tcrhoc}, we know that at this temperature, $\rhoc = A \, \mdm^4$, and thus, the relic density is given by
\beq
\rho_{\rm DM} (T_0) \,=\, A \, \mdm^4 \frac{g_*(T_0) T_0^3}{g_*(T_m) T_m^3},
\eeq{relic_1}
where $T_0$ is the current CMB temperature. Additionally, from Eq.~\leqn{rho_CFT_end}, $T_m$ is given by,
\beqa
& T_m^4  = A \, \mdm^4 \hspace{16em} \nonumber \\ 
& \times \left[  \frac{2 M_* f_d \l^2 \, g_*(T_m)}{3 (g_*(m_h))^{3/2} v }  \,\left( \frac{m_h}{\L} \right)^{2d-4}  \left( \frac{v^3}{m_h^3}-1 \right) \right]^{-1}
\eeqa{Tm}
Using Eq.~\leqn{Tm} in Eq.~\leqn{relic_1} gives the relic density of dark matter from the Higgs portal in terms of other parameters in the theory. Note that we use $g_*(T_0) \sim g_*(T_m) \sim {\cal O}(1)$. This is a reasonable approximation, as both temperatures are below the QCD scale. $g_*(m_h)$, denoted as just $g_*$ below, is approximately $\mathcal{O}(100)$. We also replace $\left( \frac{v^3}{m_h^3}-1 \right)\to {\cal O}(1)$ for this order-of-magnitude estimate. Additionally, we substitute $\gap$ in the equation instead of $\l$ and $\L$, using Eq.~\leqn{gapHH}. Taking the ratio of $\rho_{\rm DM} (T_0)$ and the present critical energy density gives Eq.~\leqn{anal_HP}:
\beqa
&&\frac{\Omega_{\scriptstyle_{\rm DM}} h^2}{0.1} = \left[ \frac{\mdm}{1\mev} \right]\left[ \frac{\left(A \, f_d^3 \, g_*^{-9/2}\right)^{1/4}}{ 10^{-5}} \right]\left[ \frac{\left( \frac{\gap}{m_h} \right)^{(6-\frac{3d}{2})}}{10^{-12}} \right]. \nonumber\\
&&\hspace{12em} 
\eeqa{relic_final}
This simple estimate is in good agreement with the results of numerical integration of Eq.~\leqn{BeqCFT}. 

\end{document}